\documentclass[a4paper]{aa}  
\usepackage{graphicx}
\usepackage{longtable}
\usepackage{txfonts}
\usepackage{ulem}
\usepackage{natbib}

\newcommand{\ltsim}{\protect\raisebox{-0.5ex}{$\:\stackrel{\textstyle <}
        {\sim}\:$}}
\newcommand{\gtsim}{\protect\raisebox{-0.5ex}{$\:\stackrel{\textstyle >}
        {\sim}\:$}}

\begin{document}
\setcounter{table}{0}
\setcounter{figure}{0}

\title{Constraining tidal dissipation in F-type main-sequence stars: \\
        the case of CoRoT-11}

    \titlerunning{Tidal dissipation in F-type stars}
    \authorrunning{A. F. Lanza et al.}


   \author{A.~F.~Lanza\inst{1} \and C.~Damiani\inst{1} \and D.~Gandolfi\inst{2}  }

   \offprints{A.~F.~Lanza}

   \institute{INAF-Osservatorio Astrofisico di Catania, Via S. Sofia, 78 
               -- 95123 Catania, Italy \\ 
              \email{nuccio.lanza@oact.inaf.it}
\and
Research and Scientific Support Department, European Space Agency, Keplerlaan 1,
 2200AG, Noordwijk, The Netherlands 
}

   \date{Received ... ; accepted ... }

\abstract{Tidal dissipation in late-type stars is presently poorly understood and the study of planetary systems hosting hot Jupiters can provide new observational constraints to test proposed theories.}{We focus on systems with F-type main-sequence stars and find that the recently discovered system CoRoT-11 is presently the best suited for such a kind of investigation.}{A classic constant tidal lag model is applied to reproduce the evolution of the system from a plausible nearly synchronous state on the ZAMS to the present state, thus putting constraints on the average modified tidal quality factor $\langle Q_{\rm s}^{\prime} \rangle$ of its F6V star. {{ Initial conditions with the stellar rotation period longer than the orbital period of the planet can be excluded on the basis of the presently observed state in which the star spins faster than the planet orbit.}}}{{{ It is found that $4 \times 10^{6} \ltsim \langle Q_{\rm s}^{\prime} \rangle \ltsim 2 \times 10^{7}$,  if the system started its evolution on the ZAMS close to synchronization, with an uncertainty related to the constant tidal lag hypothesis and the estimated stellar magnetic braking within a factor of $\approx 5-6$.}} {{ For a non-synchronous initial state of the system, $\langle Q_{\rm s}^{\prime} \rangle \ltsim 4 \times 10^{6}$ implies an age younger than $\sim 1$~Gyr, while $\langle Q_{\rm s}^{\prime} \rangle \gtsim 2 \times 10^{7}$
may be tested by comparing the theoretically derived initial orbital and stellar rotation periods with those of a sample of observed systems. Moreover,}}  we discuss how the present value of $Q_{\rm s}^{\prime}$ can be measured by a timing of the mid-epoch and duration of the transits as well as of the planetary eclipses to be observed in the infrared with an accuracy of $\sim 0.5-1$~s over a time baseline of $\sim 25$~yr. }{CoRoT-11 is an highly interesting system potentially allowing us a direct measure of the tidal dissipation in an F-type star as well as the detection of the precession of the orbital plane of the planet that provides us with an accurate upper limit for the obliquity of the stellar equator. If the planetary orbit has a significant eccentricity ($e \gtsim 0.05$), it will be possible to detect also the precession of the line of the apsides and derive information on the Love number of the planet and its tidal quality factor.}


   \maketitle


\section{Introduction}
\label{intro}
\subsection{Tidal dissipation theories}
Tidal dissipation in close binary systems with late-type components is generally constrained by the ranges of orbital periods corresponding to circular orbits as observed in clusters of different ages. \citet{OgilvieLin07} review recent observations and conclude that the equilibrium tide theory is insufficient  to explain binary circularization by at least two orders of magnitude. Therefore, in addition to the dissipation of the kinetic energy of the flow associated with the tidal bulge, which is considered in the equilibrium tide theory \citep[e.g., ][]{Zahn77,Zahn89}, other effects must be included. The dynamical tide theory treats  the dissipation of  waves excited by the oscillating tidal potential in the stellar interior whose kinetic energy is ultimately extracted from the orbital motion.  
For simplicity, we shall assume that stars are rotating rigidly. 

We consider a reference frame rotating with the stellar angular velocity $\Omega= 2\pi/P_{\rm rot}$, where $P_{\rm rot}$ is the stellar rotation period. In that frame, the tidal potential experienced by the  star can be written as a sum of rigidly rotating components proportional to the spherical harmonics 
$Y_{l m} (\theta, \phi)$, viz., ${\rm Re}\, [\Psi_{l m} r^{l} Y_{m}^{l} (\theta, \phi) \exp(-i \hat{\omega}_{lm} t) ]$, where $(r, \theta, \phi)$ are spherical polar coordinates with the origin in the centre of the star, $\hat{\omega}_{lm}$ is the tidal frequency in that frame,  $\Psi_{l m}$ the amplitude of the component of degree $l$ and azimuthal order $m$, and $t$  the time. The tidal frequency is given by $\hat{\omega}_{lm} = l n - m \Omega $, where $n = 2\pi/P_{\rm orb}$ is the  mean motion of the binary, $P_{\rm orb}$ being its orbital period.  Waves are expected to be excited with the different frequencies $\hat{\omega}_{lm}$ corresponding to the various components of the tidal potential and their amplitudes will depend on that of the exciting component and on the response of the stellar interior. For a nearly circular orbit, the $l=m=2$ component is the dominant one and is responsible for the synchronization of the stellar rotation with the orbital motion \citep[e.g., ][]{OgilvieLin04}. 

The efficiency of tidal dissipation is usually parameterized by a dimensionless quality factor $Q$ proportional to the ratio of the total kinetic energy of the tidal distortion to the energy dissipated in one tidal period $2\pi/\hat{\omega}$  \citep[e.g., ][]{Zahn08}. In the theory, $Q$ always appears in the combination $Q^{\prime} \equiv (3/2) (Q/k_{2})$, where $k_{2}$ is the Love number of the star that measures its density stratification\footnote{Note that $k_{2}$ is  twice the apsidal motion constant of the star, often indicated with the same symbol, as in, e.g., \citet{Claret95}.}. Therefore, the smaller the value of $Q^{\prime}$, the stronger the tidal dissipation. In general, $Q^{\prime}$ depends on $l$, $m$, and the tidal frequency $\hat{\omega}$, thus a rigorous treatment of the tidal dissipation should consider the sum of the effects associated with  the different tidal components each having its specific $Q^{\prime}$. In practise, we adopt a single value of $Q^{\prime}$ which represents an average of the contributions of the different components. Moreover, we also average on the tidal frequency, which means averaging  along the evolution of a given system because the tidal frequency decreases with time and goes to zero when  tidal dissipation has circularized and synchronized the binary. 

The observations reviewed by \citet{OgilvieLin07} indicate that an average $Q^{\prime}$ ranging between $5 \times 10^{5}$ and $2 \times 10^{6}$ is adequate to account for the circularization of late-type main-sequence binaries. Such  low values require an efficient tidal dissipation mechanism that  
\citet{OgilvieLin07}, moving along the lines of previous work, propose to be  the damping of  inertial waves in the stellar interior.  These waves have the Coriolis force as their restoring force and are excited  provided that the tidal frequency $\hat{\omega}$ satisfies the relationship:
\begin{equation}
|\hat{\omega} | \leq  2 \Omega. 
\label{exc_cond}
\end{equation}
The corresponding $Q^{\prime}$ has a remarkable dependence on the tidal and  stellar rotation frequencies owing to the complex details of wave excitation and dissipation that are still poorly understood \citep[cf. also][]{GoodmanLackner09}. The main point, which can be regarded as firmly established,  is that $Q^{\prime}$ decreases by $2-4$ orders of magnitude when $|\hat{\omega}|/\Omega \leq 2$ with respect to the case when 
$|\hat{\omega}|/\Omega > 2$,  because the excitation of  inertial waves is forbidden in the latter case and only the damping of the equilibrium tide contributes to the dissipation.

In view of the present uncertainties in the dynamical tide theory, a precise determination of the tidal dissipation in binary systems with well known  parameters is highly desirable. The case of F-type main-sequence stars is particularly challenging from a theoretical viewpoint because their internal structure consists of a thin outer convective zone and a radiative interior hosting a small convective core at the centre of the star. Since the propagation and dissipation of inertial waves are remarkably different in the convective and radiative zones, the study of F-type stars provides a critical test for the theory. As the mass of the outer convection zone decreases rapidly with increasing stellar mass between $1.2$ and $1.5$~M$_{\odot}$,  the value of $Q^{\prime}$ is  expected to increase by $3-4$ orders of magnitude within this mass range \citep{BarkerOgilvie09}. 

\subsection{Testing tidal theory with planetary systems}

A new opportunity to test the tidal theory comes from the  star-planet systems, in particular  those containing hot Jupiters. Systems with  transiting planets have the best determined stellar and planetary parameters and are particularly suited to study tidal dissipation { \citep[see, e.g., ][]{CaronePatzold07}}. F-type host stars having a mass $M \geq 1.2-1.5$~M$_{\odot}$ evolve quite rapidly during their main-sequence lifetime, thus improving significantly  their age estimate from model isochrone fitting in comparison with lower mass stars. A good age estimate is important to constrain the average value of $Q^{\prime}$ by modelling the tidal evolution of a particular system (cf. Sect.~\ref{back_integ}).  
 In Table~\ref{hot-Jupiters}, we list the presently known transiting systems with  a star having an effective temperature $T_{\rm eff} \geq 6250$~K which corresponds to a spectral type earlier or equal to  F8V. 

                           \begin{table*}
               \begin{tabular}{lllllllllll}
               Name     & $T_{\rm eff}$               & $M$               & $R$     & $P_{\rm rot}$     & $P_{\rm orb}$       & $M_{\rm p}$       & $R_{\rm p}$  & $\tau_{\rm syn}$          & Refs. \\
                                   & (K)     & (M$_{\odot}$)     & (R$_{\odot}$)            & (days)            & (days)     & (M$_{\rm J}$)     & (R$_{\rm J})$             & (Gyr)                  &                   \\
       & & & & & & & & & \\ 
          CoRoT-11    & 6440   $\pm $ 120    & 1.27   $\pm $ 0.05    & 1.37   $\pm $ 0.03    &  1.73   $\pm $  0.26    & 2.994    &  2.33   $\pm $ 0.34    & 1.43   $\pm $ 0.03    &   8.673    &       Ga10   \\
          HAT-P-06    & 6570   $\pm $  80    & 1.29   $\pm $ 0.06    & 1.46   $\pm $ 0.06    &  8.49   $\pm $  1.34    & 3.853    &  1.06   $\pm $ 0.12    & 1.33   $\pm $ 0.06    &  28.38    &       To08, No08   \\
          HAT-P-07    & 6350   $\pm $  80    & 1.47   $\pm $ 0.07    & 1.84   $\pm $ 0.17    & 24.51   $\pm $  5.59    & 2.205    &  1.78   $\pm $ 0.01    & 1.36   $\pm $ 0.14    &   0.166    &       Na09   \\
          HAT-P-09    & 6350   $\pm $ 150    & 1.28   $\pm $ 0.13    & 1.32   $\pm $ 0.07    &  5.61   $\pm $  0.78    & 3.923    &  0.78   $\pm $ 0.09    & 1.40   $\pm $ 0.06    & 205.8    &       Am09, Sh09   \\
  HAT-P-14   & 6600   $\pm $  90    & 1.30   $\pm $ 0.03    & 1.47   $\pm $ 0.05    &  8.85   $\pm $  0.85    & 4.628    &  2.20   $\pm $ 0.04    & 1.20   $\pm $ 0.58    &  15.21    &       To10, Si10   \\
          HAT-P-24    & 6373   $\pm $  80    & 1.19   $\pm $ 0.04    & 1.32   $\pm $ 0.07    &  6.67   $\pm $  0.68    & 3.355    &  0.69   $\pm $ 0.03    & 1.24   $\pm $ 0.07    &  58.49    &       Ki10   \\
 HD147506   & 6290   $\pm $ 110    & 1.32   $\pm $ 0.08    & 1.42   $\pm $ 0.06    &  3.14   $\pm $  0.30    & 5.633    &  8.62   $\pm $ 0.55    & 0.98   $\pm $ 0.04    &   4.379    &       Ba07, Lo08   \\
   HD15082 & 7430   $\pm $ 100    & 1.50   $\pm $ 0.03    & 1.44   $\pm $ 0.03    &  0.81   $\pm $  0.11    & 1.220    &  4.10   $\pm $ 4.00    & 1.50   $\pm $ 0.05    &   0.347    &       Ca10   \\
   HD197286   & 6400   $\pm $ 100    & 1.28   $\pm $ 0.16    & 1.24   $\pm $ 0.05    &  3.68   $\pm $  0.59    & 4.955    &  0.96   $\pm $ 0.20    & 0.93   $\pm $ 0.04    & 553.7    &       He09a   \\
         Kepler-05    & 6297   $\pm $  60    & 1.37   $\pm $ 0.06    & 1.79   $\pm $ 0.06    & 18.91   $\pm $  4.80    & 3.548    &  2.11   $\pm $ 0.06    & 1.43   $\pm $ 0.05    &   1.009    &       Ko10   \\
         Kepler-08    & 6213   $\pm $ 150    & 1.21   $\pm $ 0.07    & 1.49   $\pm $ 0.06    &  7.16   $\pm $  0.78    & 3.523    &  0.60   $\pm $ 0.19    & 1.42   $\pm $ 0.06    &  61.17    &       Je10   \\
        OGLE-TR-L9    & 6933   $\pm $  60    & 1.52   $\pm $ 0.08    & 1.53   $\pm $ 0.04    &  1.97   $\pm $  0.07    & 2.486    &  4.50   $\pm $ 1.50    & 1.61   $\pm $ 0.04    &   0.927    &       Sn09   \\
           WASP-03    & 6400   $\pm $ 100    & 1.24   $\pm $ 0.09    & 1.31   $\pm $ 0.09    &  4.95   $\pm $  0.91    & 1.847    &  1.76   $\pm $ 0.09    & 1.29   $\pm $ 0.09    &   1.002    &       Gi08, Po08   \\
           WASP-12    & 6250   $\pm $ 150    & 1.35   $\pm $ 0.14    & 1.57   $\pm $ 0.07    & 36.12   $\pm $ 49.03    & 1.091    &  1.41   $\pm $ 0.09    & 1.79   $\pm $ 0.09    &   0.012    &       He09   \\
           WASP-14    & 6475   $\pm $ 100    & 1.21   $\pm $ 0.12    & 1.31   $\pm $ 0.07    & 23.68   $\pm $  6.35    & 2.244    &  7.34   $\pm $ 0.50    & 1.28   $\pm $ 0.08    &   0.017    &       Jos08, Jo09   \\
           WASP-15    & 6300   $\pm $ 100    & 1.18   $\pm $ 0.12    & 1.48   $\pm $ 0.07    & 18.69   $\pm $ 13.64    & 3.752    &  0.54   $\pm $ 0.05    & 1.43   $\pm $ 0.08    &  22.41    &       We09   \\
           WASP-17    & 6550   $\pm $ 100    & 1.20   $\pm $ 0.12    & 1.38   $\pm $ 0.20    &  7.76   $\pm $  2.49    & 3.735    &  0.49   $\pm $ 0.06    & 1.74   $\pm $ 0.24    & 128.7    &       An09   \\
           WASP-18    & 6400   $\pm $ 100    & 1.25   $\pm $ 0.13    & 1.22   $\pm $ 0.07    &  5.60   $\pm $  1.11    & 0.941    & 10.30   $\pm $ 0.69    & 1.11   $\pm $ 0.06    &   0.002    &      He09b   \\
              XO-3    & 6429   $\pm $ 100    & 1.21   $\pm $ 0.07    & 1.38   $\pm $ 0.08    &  3.81   $\pm $  0.23    & 3.192    & 11.79   $\pm $ 0.59    & 1.22   $\pm $ 0.07    &   0.814    &       Joh08, Wi09   \\
              XO-4    & 6397   $\pm $  70    & 1.32   $\pm $ 0.02    & 1.56   $\pm $ 0.05    &  8.97   $\pm $  0.80    & 4.125    &  1.78   $\pm $ 0.08    & 1.34   $\pm $ 0.05    &  11.32    &       Mc08, Na10   \\
               \end{tabular}
	\caption{Parameters of the  transiting planetary systems having stars with $T_{\rm eff} \geq 6250$~K;  M$_{\rm J}=1.90 \times 10^{27}$~kg and R$_{\rm J}=7.15 \times 10^{7}$~m indicate the mass and the radius of Jupiter. {References}:  Am09: \cite{Ammleretal09}; An09: \cite{Andersonetal10}; Ba07: \cite{Bakosetal07}; Ca10: \cite{Cameronetal10}; Ga08: \cite{Gandolfietal10}; Gi08: \cite{Gibsonetal08}; He09a : \cite{hellieretal09a}; He09b: \cite {hellieretal09b}; He09: \cite{Hebbetal09}; Je10: \cite{Jenkinsetal10}; Jos08: \cite{Joshietal08}; Joh08: \cite{JohnsKrulletal08}; Jo09: \cite{Johnsonetal09}; Ki10: \cite{Kippingetal10}; Ko10 : \cite{Kochetal10}; Lo08: \cite{Loeilletetal08};  Mc08: \cite{McCulloughetal08}; Na09: \cite{Naritaetal09}; Na10: \cite{Naritaetal10}; No08: \cite{Noyesetal08}; Po08: \cite{Pollaccoetal08}; Sh09: \cite{Shporeretal09}; Si10: \cite{Simpsonetal10}; Sn09: \cite{Snellenetal09}; To08 : \cite{Torresetal08}; To10: \cite{Torresetal10};  We09: \cite{Wessetal09}; Wi09: \cite{Winnetal09b}}

	\label{hot-Jupiters}
                            \end{table*}


The columns from left to right list the name of the system, the effective temperature $T_{\rm eff}$, the mass $M$ and the radius $R$  of the star, its rotation period $P_{\rm rot}$, as derived from the observed spectroscopic rotation broadening $v \sin i$  and the estimated stellar radius { assuming a equator-on view of the star}, the orbital period  $P_{\rm orb}$, the mass $M_{\rm p}$ and radius $R_{\rm p}$ of the planet, 
 the  timescale for the synchronization of the stellar rotation $\tau_{\rm syn}$, and the references. To compute the synchronization time, we  assume that the entire star is synchronized as is customary in tidal theory and is suggested by the  tidal evolution of close binaries  observed in stellar clusters of different ages. The synchronization timescale is then a measure of the strength of  tidal dissipation in the star and is computed  according to the formula: 
\begin{equation}
\tau_{\rm syn}^{-1} \equiv \frac{1}{\Omega} \left| \frac{d\Omega}{dt} \right| = \frac{9}{2} \frac{1}{\gamma^{2} Q^{\prime}_{\rm s}} \left( \frac{M_{\rm p}}{M} \right)^{2} \left( \frac{R}{a} \right)^{9/2} \left| 1 - \left( \frac{n}{\Omega} \right) \right| \sqrt{\frac{G M}{R^{3}}},
\label{tidal_time}
\end{equation}
where $\gamma R \simeq 0.22 \, R$ is the gyration radius of the star \citep{Siessetal00}, $Q_{\rm s}^{\prime}$ its modified tidal quality factor, here assumed to be $Q^{\prime}_{\rm s} = 10^{6}$, $a$ the semimajor axis of the orbit,  and $G$ the gravitation constant \citep[see ][]{MardlingLin02}. Equation~(\ref{tidal_time}) is valid for circular orbits and when the spin axis is aligned with the orbital angular momentum. In this regard, the values given here must be considered as estimations and for illustration purpose only, as high eccentricity and/or obliquity have been measured for some of these systems. 

According to the dynamical tide theory by \citet{OgilvieLin07}, the stars experiencing the strongest tidal interaction are those with $n/\Omega = P_{\rm rot}/P_{\rm orb} \leq 2$ because they have $|\hat{\omega}|/\Omega \leq 2$ when the $l=m=2$ component of the tidal potential is considered \citep{BarkerOgilvie09}. For such stars, the orbital decay (or expansion) due to the tidal interaction can in principle be observed. Among those systems, the best candidate is CoRoT-11 \citep{Gandolfietal10} because it has a tidal synchronization timescale in between those derived for systems like HD~197286/WASP-7 or HAT-P-09, i.e., much longer than the main-sequence lifetime of the system, and those of, e.g., OGLE-TR-9 or WASP-3 that are shorter than the expected ages of the systems, implying that the  synchronous final state has possibly already been reached. Other systems, e.g., HD15082/WASP-33,  have a star so massive that a  $Q^{\prime}_{\rm s}$ too large to be measurable is expected, or show a remarkable misalignment between the stellar spin and the orbital angular momentum that makes the derivation of the rotation period from the $v \sin i$ quite uncertain, as in the case of XO-4. 

In view of the peculiar characteristics of { the CoRoT-11 system}, we shall consider it for a detailed study of the tidal evolution. We shall derive constraints on the average $Q^{\prime}_{\rm s}$ value of its F6V star by considering a possible initial state for the system when its star settled on the zero-age main-sequence (hereafter ZAMS; see Sect.~\ref{initial_status}). Moreover, we shall demonstrate how the present value of $Q^{\prime}_{\rm s}$ can be directly measured with suitable transit observations extended on a time interval of a few decades.

\section{Method: tidal evolution model and initial conditions}
\label{tidal_model}
\subsection{The constant quality factor approximation}
We adopt the classic equilibrium tidal model after \citet{Hut81}, in the formulation given by \citet{Leconteetal10}. It assumes that the energy of the tidal perturbation raised by the tidal potential is dissipated by viscous effects  producing a {\it constant} time lag $\Delta t$ between the maximum of the tidal potential and the tidal bulge. When $Q \gg 1$, $Q^{-1} \simeq \hat{\omega} \Delta t \sim 2 \delta(\hat{\omega})$, where $\delta(\hat{\omega})$ is the  lag angle between the maximum of the deforming potential and the tidal bulge. Since $\Delta t$ is assumed to be constant during the tidal evolution, $Q$ and $Q^{\prime}$ vary as $\hat{\omega}^{-1}$. To model the evolution of a system by assuming a constant (or,  better, an average) $Q^{\prime}$, 
we need to make an approximation. Following \citet{Leconteetal10}, we assume for the star and the planet 
$k_{\rm 2s} \Delta t_{\rm s} = 3/(2n_{\rm obs} Q^{\prime}_{\rm s})$ and $k_{\rm 2p} \Delta t_{\rm p} = 3/(2n_{\rm obs} Q^{\prime}_{\rm p})$,  where the indexes s and p refer to the star and the planet, respectively, 
and $n_{\rm obs}$ is the present mean orbital motion of the system corresponding to the orbital period $P_{\rm orb}=2.99433$~days. As we shall see (cf. Sect.~\ref{applications}), the mean motion $n$ varies by a factor of $\approx 4-5$ during the evolution of the system, but we shall neglect such a change because we are interested in deriving the order of magnitude of the average $Q^{\prime}_{\rm s}$ along the evolution. 

The theory of dynamical tides predicts that the dissipation is  a complex function of the tidal frequency, the properties of the interior of the body, and, possibly, the strength of the tidal potential because of non-linear effects. All such dependencies are lumped together  into the coefficients $Q^{\prime}_{\rm s}$ and 
$Q^{\prime}_{\rm p}$ and in principle could be included in an equilibrium model provided that the relationship between their variations along the evolution of the system and the  time lag value were known. In view of our ignorance of the processes contributing to the tidal  dissipation in stars and planets \citep[cf., e.g.,][]{OgilvieLin04,GoodmanLackner09} and the presently limited observational constraints, we shall consider  constant, i.e. average, values of $Q^{\prime}_{\rm s}$ and $Q_{\rm p}^{\prime}$ all along the evolution and integrate the equations in Sect.~2.2 of \citet{Leconteetal10} accordingly.

\subsection{Range of { considered} $Q^{\prime}_{\rm s}$}
\label{Qstar}
The dynamical tide theory of \citet{BarkerOgilvie09} gives  the dependence of $Q^{\prime}_{\rm s}$ on the tidal and the rotation frequencies of the star when the dissipation of inertial waves is regarded as the main source of   energy dissipation -- see their Fig.~8, and consider that $Q^{\prime}_{\rm s} \propto \Omega^{-2}$ for a fixed ratio $\hat{\omega}/\Omega$ \citep[cf. Sect.~3.6 of ][]{OgilvieLin07}. In view of the simplified treatment of the interaction of such waves  with turbulent convection and radiative-convective boundaries, their results can be regarded only as an approximation of the complex dependence of $Q^{\prime}_{\rm s}$ on the relevant parameters. For a star of 
$1.2$ M$_{\odot}$ and $P_{\rm rot} \sim 1.5$~days, the average $Q^{\prime}_{\rm s}$ for $|\hat{\omega}|/\Omega \leq 2$ can be roughly estimated to be $ 5 \times 10^{7}$ with an uncertainty of at least one order of magnitude and some preference for the lower bound because of the approximate treatment of the above mentioned processes. Outside such a frequency range, $Q^{\prime}_{\rm s} \sim  10^{10}$, i.e., the tidal dissipation is reduced by at least three orders of magnitudes. 

\subsection{Range of { considered} $Q^{\prime}_{\rm p}$}
{ An estimate of the tidal quality factor of the hot Jupiters may be based on astrometric observations of the Jupiter-Io system that give $  Q^{\prime}_{\rm p} \simeq (1.36 \pm 0.21) \times 10^{5}$ for Jupiter \citep[see ][]{Laineyetal09}}. Nevertheless, the quality factor of hot Jupiters can be different from that of Jupiter owing to their different internal structure. Assuming the observed eccentrity of the orbits of several hot Jupiters  to be of primordial origin, \citet{Matsumuraetal08} infer $10^{5} \leq Q_{\rm p}^{\prime} \leq 10^{9}$, with most of the planets having $Q^{\prime}_{\rm p} \leq 10^{8}$. Therefore, we shall consider $ 10^{6} \leq Q^{\prime}_{\rm p} \leq 10^{8}$ when integrating the equation of tidal evolution of our system. The timescale to achieve the pseudosynchronization of the planetary rotation in the case of CoRoT-11 and of the other systems containing hot Jupiters is of the order of $10^{5}$ yr when $Q^{\prime}_{\rm p} \sim 10^{6}$ \citep[cf. Sect. 2.2 of  ][]{Leconteetal10}. Even if $Q^{\prime}_{\rm p} \sim 10^{8}$ such a timescale is shorter than $3 \times 10^{7}$ yr, i.e., much shorter than the typical timescales for the tidal evolution of the orbital parameters and the stellar spin.
Therefore, we shall consider the planet to be always in a state of pseudosynchronization which simplifies the integration of the  tidal equations.

\subsection{Initial state of the tidal evolution of planetary systems}
\label{initial_status}

\begin{figure}[t]
\centering
\includegraphics[width=1.0\linewidth]{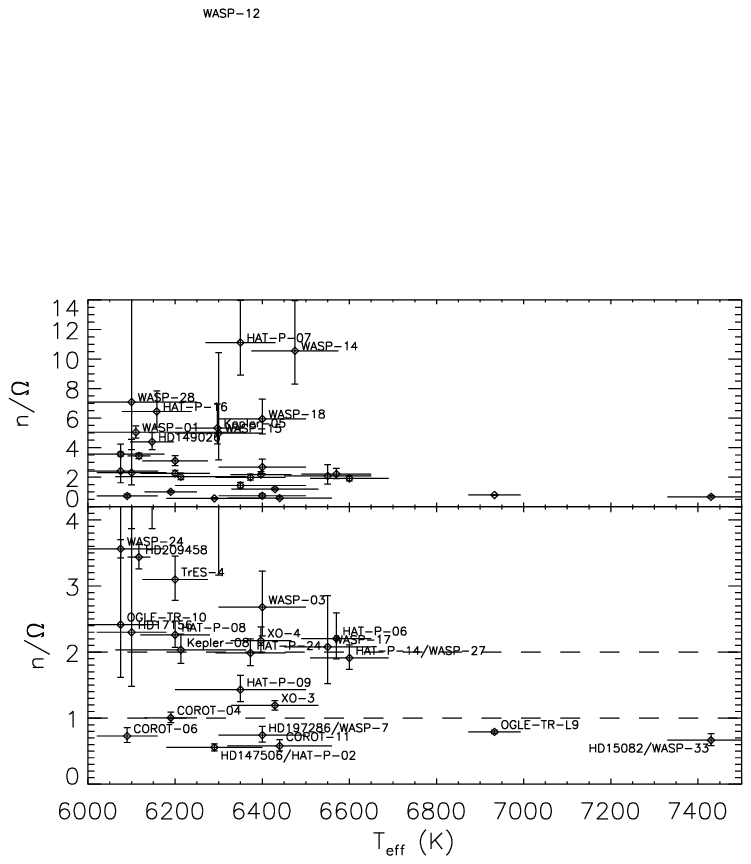} 
\caption{Upper panel: The synchronization parameter $n/\Omega$ vs. the effective temperature of the star $T_{\rm eff}$ in transiting  planetary systems with  $T_{\rm eff} \geq 6000$ K. WASP-12 with $n/\Omega \approx$ 33 has been omitted for the sake of clarity; lower panel: an enlargement of the lower portion of the upper panel, to better show the domains close to  $n/\Omega=1$ and  $n/\Omega = 2$, which are marked by horizontal dashed lines. The names of the systems are reported in both panels, although they are omitted for $n / \Omega \leq 4.0$ in the upper panel to avoid confusion.  
 }
\label{noveromega}
\end{figure}

The observed distribution of the synchronization parameter $n/\Omega \equiv P_{\rm rot}/P_{\rm orb}$ provides us with information on the initial state of star-planet systems. We show $n/\Omega$ vs. the effective temperature { $T_{\rm eff}$} of the host star for stars with $T_{\rm eff} \geq 6000$ K in Fig.~\ref{noveromega} where the most impressive feature is the lack of systems in the range $1 \leq n/\Omega \leq 2$, with the exception of HAT-P-09 and XO-3.

 As conjectured by \citet{Lanza10}, this may be related to the processes occuring in the systems before their stars settle on the ZAMS. After the planet forms in the circumstellar disc, it migrates inwards by transferring its orbital angular momentum to the disc via resonant interactions \citep[e.g., ][]{Linetal96}. The migration ends where the orbital period of the planet is half the orbital period of the particles at the inner edge of the disc where it is truncated by the magnetic field of the protostar. Since the protostellar magnetic field is also responsible for the locking of the stellar rotation to the inner edge of the disc, the value of the synchronization ratio at this stage is $n/\Omega \equiv P_{\rm rot}/P_{\rm orb} \simeq 2$. After a few million years, the disc disappears and the subsequent evolution depends on the interplay between tidal and magnetic interactions coupling the star and the planet. As the star contracts towards the ZAMS, it accelerates its rotation leading  to a decrease of $n/\Omega$ \citep[cf. Fig.~8 of ][]{Lanza10}. The loss of angular momentum through the stellar magnetized wind, however, counteracts such an acceleration. If the lifetime of the disc is $\ltsim 10-15$~Myr, the contraction prevails and the system enters into the domain $n/\Omega< 2$ characterized by a strong tidal interaction due to the dissipation of inertial waves in the extended convective envelope of the pre-main-sequence star \citep{OgilvieLin07}. Angular momentum is transferred from the orbit to the spin of the star, leading to the infall of the planet towards the star \citep{Hut80}. Considering a star similar to the Sun, $Q_{\rm s}^{\prime} = 10^{5}$, a planet of $M_{\rm p} = 1$~M$_{\rm J}$, and an orbital period of 3 days,  eq.~(8) of \citet{OgilvieLin07} gives an infall timescale of $\approx 250$~Myr, that may account for the lack of systems with $1 < n/\Omega < 2$ given that typical stellar ages are between $\sim 1$ and $\sim 8$~Gyr.

On the other hand, if the magnetic braking  prevails, the system may be driven into the domain where $n/\Omega > 2$ and the tidal interaction becomes so weak ($Q_{\rm s}^{\prime} \sim 10^{8}-10^{9}$) that the planet will spiral towards the star on a timescale much longer than the main-sequence lifetime of the star \citep{OgilvieLin07}, thus populating the region of the diagram where $n/\Omega >2$. { Only when the planet is more massive than $\sim 5-10$~M$_{\rm J}$ and the star is  late~F or cooler, the tidal interaction may lead to the infall of the planet during the main-sequence lifetime of the star \citep[cf., e.g., ][]{Bouchyetal11}.}

Another possible scenario occurs when the magnetic field of the pre-main-sequence star is so strong ($\approx 3-5$~kG) as to effectively couple the orbital motion of the planet to the rotation of the star after the disc has disappeared  \citep{Lovelaceetal08,Vidottoetal10}. In this case, a synchronous state is attained on a timescale of $\approx 2-3$~Myr and maintained until the field strength decreases rapidly as the star gets close to the ZAMS. If the magnetic braking prevails over the final stellar contraction, the system would reach the ZAMS with  \textbf{$1 < n/\Omega < 2$} and the planet will eventually fall into the star owing to the strong tidal interaction. On the contrary, if the stellar contraction prevails, the star will spin up, reach $n/\Omega < 1$ on the ZAMS, and the tides will push the planet away from the star. This may explain why several systems populate the domain $n / \Omega < 1$ in the diagram. The present value of $n/\Omega$ is smaller than the initial value on the ZAMS because the tidal interaction transfers angular momentum from the stellar spin to the planetary orbit leading to an increase of the semimajor axis. Thus we shall base our calculations on the hypothesis that the systems observed today with $n / \Omega \leq 1 $, as  CoRoT-11,  have reached the ZAMS in a state close to synchronization. 

Finally, remarks may be done concerning the two systems that have $1 < n/\Omega < 2$. The peculiar state of the HAT-P-9 system having $n / \Omega \simeq 1.4$ may be explained by the weakness of the tidal interaction owing  to the relatively small planetary mass ($\sim 0.7$ M$_{\rm J}$), and/or the young age of the system. For XO-3, the system has significant measured eccentricity and projected obliquity ($e=0.288$ and $\lambda = 37.3\pm 3.7^{\circ}$ respectively), so it is possible that the rotation period of the star is ill estimated or that the system has undergone a different evolution.

\section{Application to CoRoT-11}
\label{applications}

\subsection{System parameters}
\citet{Gandolfietal10} present the observations leading to the discovery of CoRoT-11 and derive the parameters of the system that are reported in their Table~4 together with their uncertainties. The stellar rotation period can be inferred only from the observed $v \sin i $ and the estimated radius of the star because no evidence of rotational modulation of the optical flux has been found in  CoRoT photometry. As a matter of fact, a modulation with a period of $\approx 8$ days  appears in the photometric time series but it is incompatible with the maximum rotation period of $ 1.73 \pm 0.25$ days as derived  from the $v \sin i = 40.0 \pm 5$~km~s$^{-1}$ and the star radius 
$R= 1.37 \pm 0.03$ R$_{\odot}$. It is likely to arise from  a nearby contaminant star about 2 arcsec from CoRoT-11 that falls inside the CoRoT photometric mask. 

The eccentricity of the planetary orbit is ill-constrained by the radial velocity curve due to the uncertainty of the radial velocity measurements for such a rapidly rotating F-type star. An upper limit of $0.2$ comes from the modelling of the transit light curve that would lead to a mean stellar density incompatible with a F6V star for greater values. In any case, the present radial velocity and transit data are fully compatible with a circular orbit.

The angle $\lambda$ between the projections of the stellar spin and the orbital angular momentum on the plane of the sky can be measured through the Rossiter-McLaughlin effect \citep[e.g., ][]{Ohtaetal05}. Owing to the limited precision of the available measurements, \citet{Gandolfietal10} could not measure  $\lambda$, but they established that the planet orbits in a prograde direction, i.e., in the same direction as the stellar rotation, and that $\lambda \ltsim 50^{\circ}-80^{\circ}$. An aligned system, i.e., having $\lambda = 0^{\circ}$,  is  compatible with their observations. The obliquity $\epsilon$ of the stellar equator with respect to the orbital plane of the planet is given by:
$\cos \epsilon = \cos I_{\rm p} \cos i + \sin I_{\rm p} \sin i \cos \lambda$, where $I_{\rm p}$ and $i$ are the angles of inclination of the  orbital plane and the stellar equator to the plane of the sky. Since $I_{\rm p} = 83\fdg3$, $\cos \epsilon \simeq \sin i \cos \lambda$, thus the measurement of $\lambda$ provides us with a lower limit for the obliquity of the stellar rotation axis. 

\subsection{Stellar magnetic braking}
{ Late-type stars lose angular momentum through their magnetized stellar winds. In the case of F-type stars, 
\citet{BarkerOgilvie09} assumed a braking law of the form: $(d \Omega / d t)_{\rm mb} = - \gamma_{\rm w} \Omega^{3}$, where $\gamma_{\rm w} = 1.5 \times 10^{-15}$~yr is a factor that accounts for the efficiency of  magnetic braking. It can be integrated to give: $\Omega (t) = \Omega_{0} [1 + \gamma_{\rm w} \Omega_{0}(t-t_{0})]^{-0.5}$, where $\Omega_{0}$ is the angular velocity at initial time $t_{0}$. For $\Omega \ll \Omega_{0}$ this yields the usual Skumanich braking law of stellar rotation. 
Applying that equation to CoRoT-11a leads to an unrealistically fast initial rotation for an age between 1 and 3~Gyr. As a matter of fact, the above law of angular momentum loss does not apply to fast rotators such as CoRoT-11a for which a saturation of the angular momentum loss rate is required to explain the observed distribution of stellar rotation periods \citep[cf., e.g. ][]{Bouvieretal97}. Therefore, we shall assume:
 $(d \Omega / d t)_{\rm mb} = - \gamma_{\rm w} \Omega_{\rm sat}^{2} \Omega $, where $\Omega_{\rm sat}$ is the angular velocity corresponding to the transition to the unsaturated regime. The value of $\Omega_{\rm sat}$ for solar-mass stars is $\approx 8 \Omega_{\odot}$, where $\Omega_{\odot}$ is the angular velocity of the present Sun 
\citep{IrwinBouvier09}. However, its value is highly uncertain in the case of mid-F stars because they do not show a rotational modulation measurable from the ground and experience a remarkably lower magnetic braking than G-type  stars. Considering the data in Table~4 of \citet{WolffSimon97}, we estimate a rotation period corresponding to  saturation of $\approx 2.8$ days for mid-F stars with  $v \sin i \sim 20-25$~km~s$^{-1}$ at the age of the Hyades and the adopted value of $\gamma_{\rm w}$.  However, if CoRoT-11a is  older than $\sim 2$~Gyr, the  value of the saturation period must be  longer by a factor of at least  $\approx 2$ otherwise the star would have been rotating faster than the brake-up velocity on the ZAMS. This reduction of the angular momentum loss rate in the saturated regime can be a consequence of the initially fast angular velocity that led to a supersaturated dynamo regime for CoRoT-11a, or an effect of the close-in massive planet that reduced the efficiency of the stellar wind to extract angular momentum from the star \citep[cf. ][]{Lanza10}. 
}

\subsection{Backward integration from the present state to the initial condition}
\label{back_integ}

\begin{figure}[t]
\begin{center}
\includegraphics[angle=0, width=1.0\columnwidth]{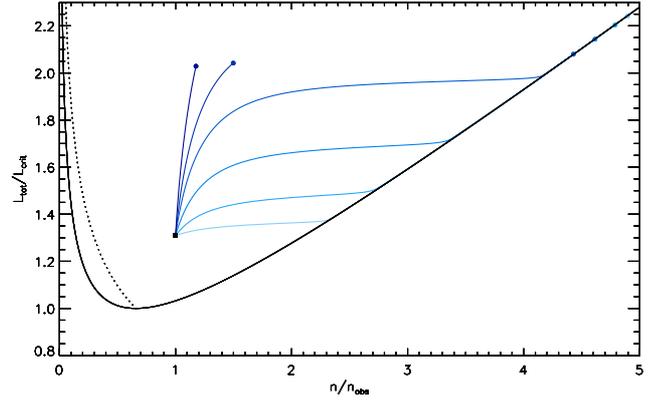}
\caption{The total angular momentum of the CoRoT-11 system in units of the critical angular momentum as defined in Eq.~(4.1) of \citet{Hut80} versus the orbital mean motion in units of the present mean motion $n_{\rm obs}$. The stellar magnetic braking law has $P_{\rm sat}=6.0$~days. The black solid line indicates the total angular momentum values corresponding to tidal equilibrium. The black dotted line represent the stability limit for the equilibrium as given by Eq.~(4.2) of \citet{Hut80}. The solid blue lines show the backwards evolution of the system for different values of the tidal quality factor $Q_{\rm s}^{\prime}$.  
From left to right, the values of $Q^{\prime}_{\rm s}$ are: $5 \times 10^7$, $3 \times 10^7$, $ 2 \times 10^7$, $10^7$, $ 3.8 \times 10^6$, and  $10^6$, respectively. The initial points of the evolution are marked by the filled blue dots. }
\label{total_ang_mom}
\end{center}
\end{figure}

\begin{figure}[t]
\begin{center}
\includegraphics[angle=0, width=1.0\columnwidth]{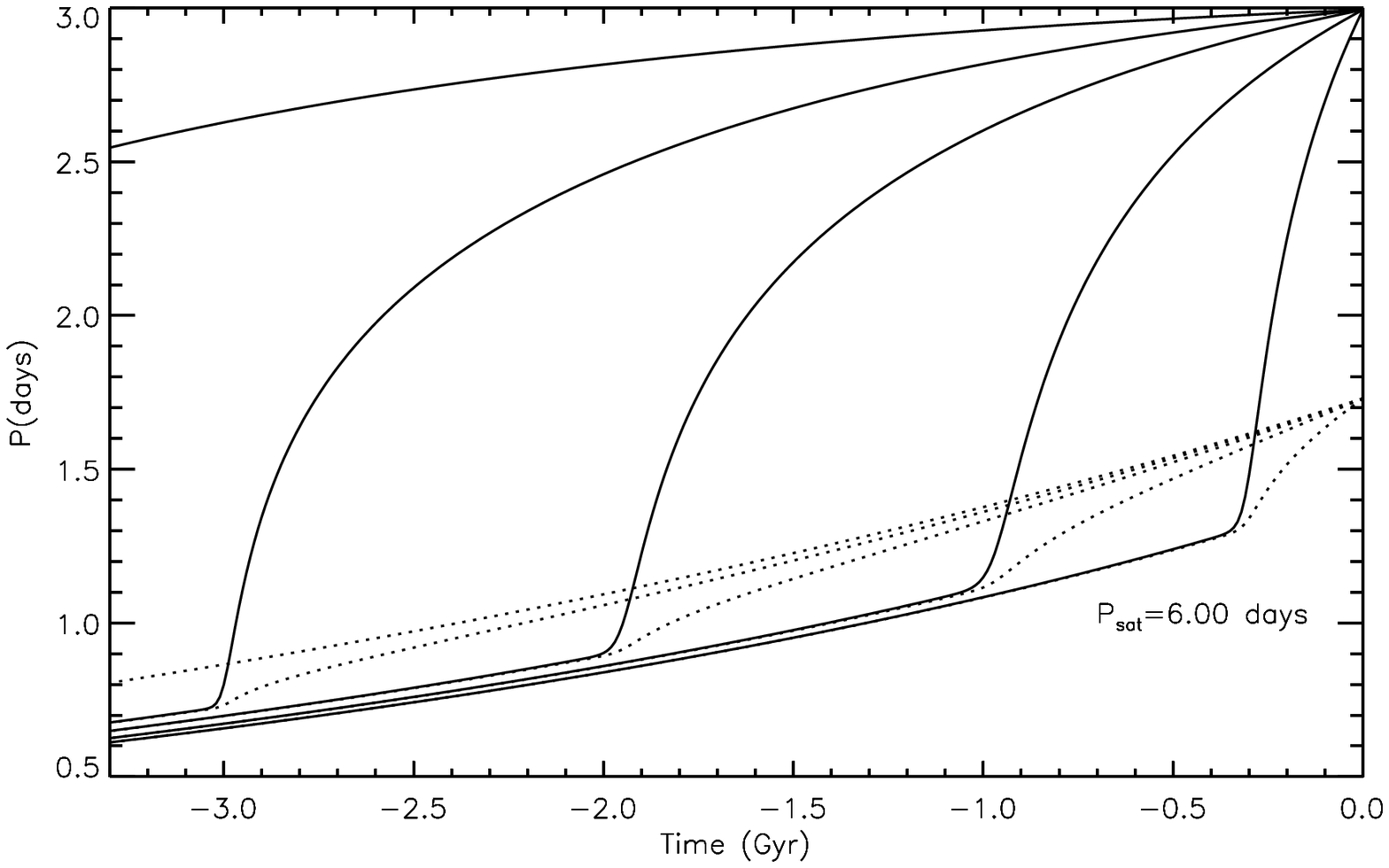}
\caption{The backward evolution of the orbital period $P_{\rm orb}$ (solid line) and the rotation period $P_{\rm rot}$ (dotted line) vs. time since the present epoch for different values of the modified tidal quality factor $Q^{\prime}_{\rm s}$ and a stellar braking law with $P_{\rm sat}=6.0$~days. From left to right, the values of $Q^{\prime}_{\rm s}$ are: $5 \times 10^7$, $2 \times 10^7$, $ 10^7$, $ 3.8 \times 10^6$, and  $10^6$, respectively. }
\label{reverse_evol}
\end{center}
\end{figure}
\begin{figure}[t]
\begin{center}
\includegraphics[angle=0, width=1.0\columnwidth]{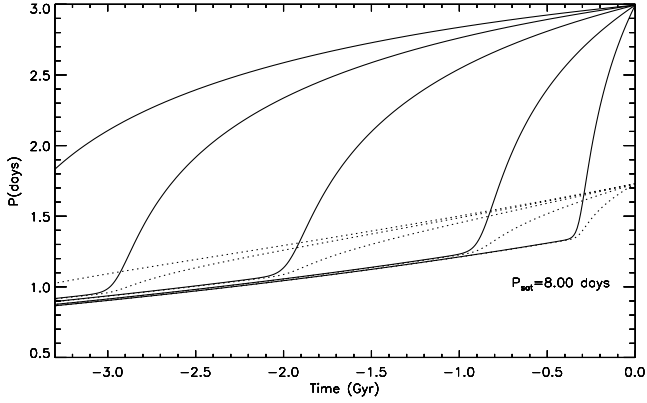} 
\caption{The backward evolution of the orbital period $P_{\rm orb}$ (solid line) and the rotation period $P_{\rm rot}$ (dotted line) vs. time since the present epoch for different values of the modified tidal quality factor $Q^{\prime}_{\rm s}$ and a stellar braking law with $P_{\rm sat}=8.0$~days. From left to right, the values of $Q^{\prime}_{\rm s}$ are: $2 \times 10^7$, $1.4 \times 10^7$, $ 8 \times 10^6$, $ 3 \times 10^6$, and  $10^6$, respectively. }
\label{reverse_evol1}
\end{center}
\end{figure}

We integrate the tidal evolution equations in Sect.~2.2 of \citet{Leconteetal10} backwards in time starting from the present condition, i.e., with the following parameters: stellar rotation period $P_{\rm rot}=1.73$ days, 
planet orbital period $P_{\rm orb} = 2.99433$ days, star mass $M=1.27$ M$_{\odot}$, star radius $R= 1.37$ R$_{\odot}$, planet mass $M_{\rm p} = 2.33$~M$_{\rm J}$, planet radius $R_{\rm p} = 1.43$~R$_{\rm J}$ (where M$_{\rm J}$ and R$_{\rm J}$ are the mass and radius of Jupiter), orbital eccentricity $e=0$, and stellar obliquity $\epsilon =0^{\circ}$. The  stellar quality factor $Q^{\prime}_{\rm s}$ is regarded as a free parameter (see also  Sect. \ref{Qstar}). { We account for the angular momentum loss of the star due to its magnetized wind by adopting a saturated  loss rate with two possible values for $P_{\rm sat}=2 \pi/\Omega_{\rm sat}$, i.e., $6.0$ or $8.0$ days.} Note that the results do not depend on the planet quality factor $Q^{\prime}_{\rm p}$, provided that the orbital eccentricity is zero and the planet is assumed to be in the synchronous rotation state.

{ The present state of the system and its backwards evolution are represented in a tidal stability diagram similar to that of \cite{Hut80} in Fig.~\ref{total_ang_mom}. The black solid line represents the value of the total angular momentum of the system when  tidal equilibrium is established. For each value of the total angular momentum exceeding a minimum critical value \citep[see ][]{Hut80}, two equilibrium states are possible. The one on the left of the dotted line is stable,  the other on the right is unstable.
The filled square indicates the present state of the CoRoT-11 system with the assumption that $i =90^{\circ}$, i.e., the stellar rotation period is 1.73~days.

  The results of the backwards integrations of the tidal equations are represented by the solid blue lines for different values of $Q_{\rm s}^{\prime}$ and a braking law with $P_{\rm sat}=6.0$~days. All the backwards evolutions asymptotically approach an initial equilibrium state where all the time derivatives of the relevant variables vanish. Such an equilibrium is unstable, i.e., it leads to the transfer of angular momentum from the stellar spin to the planet that is pushed outwards. Increasing the value of $Q_{\rm s}^{\prime}$ makes the tidal interaction weaker so that the evolution of the total angular momentum is dominated by the magnetic braking of the star. This  leads to an almost vertical evolution track in the diagram since the star was initially rotating faster. There is no  possible evolution track that crosses the dashed line, indicating that the past system evolution has occured only in the unstable region of the diagram, i.e., the planet has been pushed outwards starting from its initial orbit with $n/\Omega < 1$ during all the system lifetime. 

Given that the initial state is unstable, integrating the tidal evolution equations backwards in time is recommended to ensure the stability of the obtained solutions. The instability of the present tidal state is not a specific property of CoRoT-11, but it is generally   found in the case of hot Jupiter systems \citep{Levrardetal09} most of which evolve towards the infall of the planet into the star. Therefore, what makes CoRoT-11 peculiar is the small value of the $n/\Omega$ ratio pushing the planet outwards.} 

{ The  orbital period of the planet and the rotation
period of the star are plotted vs.  time in Fig.~\ref{reverse_evol} for $P_{\rm sat}=6.0$ days. 
If the initial state of the system was characterized by a synchronization of the stellar rotation with the planet orbital period,  it is possible to constrain the tidal quality factor, i.e., $3.8 \times 10^{6} \leq Q_{\rm s}^{\prime} \leq 2 \times 10^{7}$. Although we speculated that the system may have reached the ZAMS close to synchronization in Sect.~\ref{initial_status}, other evolutionary scenaria are possible. In some of these scenaria, the planet may have started with an orbital period longer than the stellar rotation period 3~Gyr ago, leading to the present state if $Q_{\rm s}^{\prime} \geq 2 \times 10^{7}$. As an example of such an evolution, we plot in Fig.~\ref{reverse_evol} the case of $Q_{\rm s}^{\prime} = 5 \times 10^{7}$. Note that in this case the initial state 
of the system has $n/\Omega \simeq 0.25$ that is significanly smaller than the observed values of $n/\Omega$ plotted in Fig.~\ref{noveromega}. Although the present sample of systems with $n/\Omega < 1$ is too small to draw conclusions on the minimum value of $n/\Omega$ during their evolution, we hope that such a statistical information can be used to constrain $Q_{\rm s}^{\prime}$ when a larger sample is available. 

On the other hand, $Q_{\rm s}^{\prime} < 3.8 \times 10^{6}$ implies a system younger than $\sim 1$~Gyr. To illustrate this point, we plot in Fig.~\ref{reverse_evol} the evolution for $Q_{\rm s}^{\prime} = 10^{6}$ for which the system reached synchronization $\sim 0.4$~Gyr ago. Although the backwards evolution can be continued to earlier times, this initial state is unstable as well as all the allowed initial states with $P_{\rm orb} > P_{\rm rot}$. For $Q_{\rm s}^{\prime} < 3.8 \times 10^{6}$, the tidal interaction would be so strong and the evolution   so fast that the system would reach a present value of $P_{\rm orb}$ longer than observed.  Therefore, we can set a lower limit on $Q_{\rm s}^{\prime}$ on the basis of the estimated age of the system. Note that if the system were younger than $\sim 0.7-1$~Gyr, the $Q^{\prime}_{\rm s}$ value could be so small that we may directly measure it after a few decades of timing observations (see Sect.~\ref{present_diss}).

The evolution tracks obtained for $P_{\rm sat}=8.0$~days and different values of the average $Q_{\rm s}^{\prime}$ are plotted in Fig.~\ref{reverse_evol1}. They show that a decrease of the braking rate of CoRoT-11a by $\sim 30$ percent leads to a comparable decrease in the values of the average $Q_{\rm s}^{\prime}$ because a stronger tidal interaction is needed to transfer the required amount of angular momentum from the stellar spin to the planet when the stellar rotation was initially slower. On the other hand, if we adopt $P_{\rm sat} = 2.8$~days, as estimated from the data of \citet{WolffSimon97}, the stellar braking was so rapid that only in the case of a system younger than 1~Gyr we may find an acceptable evolution with $Q_{\rm s}^{\prime} \sim 9\times 10^{6}$. For higher values of $Q_{\rm s}^{\prime}$, the evolution takes longer and this  leads to a planet approaching the Roche lobe radius and/or an initial stellar rotation exceeding the break-up velocity. 
Therefore, the assumption that $P_{\rm sat}$ is at least $\sim 5-6$~days seems to be more likely for CoRoT-11a than $P_{\rm sat} \sim 3-4$ days. In the latter case, the system would be so young and the average $Q_{\rm s}^{\prime}$ so small that we expect  to detect the tidal increase of the orbital period within a few decades  (see Sect.~\ref{present_diss}). 
}

\subsubsection{Initial eccentricity of the orbit}

In the  scenario introduced in Sect.~\ref{initial_status}, the initial eccentricity of the planetary orbit would be close to zero \citep{SariGoldreich04},  unless an encounter with some planetary or stellar body would have  excited it. Assuming that no third body has been perturbing the system after that event, the evolution of any primordial  eccentricity is ruled by the dissipation inside both the star and the planet, the latter being the most important for $Q_{\rm p}^{\prime} \ltsim 10^{8}$. The timescale for the damping of the eccentricity $\tau_{\rm e}^{-1} \equiv (1/e) (de/dt)$ adopting the present parameters of the system is $\sim 120$ Myr for $e=0.2$, $Q_{\rm s}^{\prime} = 5 \times 10^{6}$, and $Q_{\rm p}^{\prime} = 10^{6}$,  and increases to $\sim 1.1$ Gyr for $Q_{\rm p}^{\prime} = 10^{7}$. Since the tidal interaction was initially stronger -- it scales as $a^{-5}$ -- the initial $\tau_{\rm e}$ was one order of magnitude shorter than the above values, leading to a rapid damping of any initial eccentricity, unless $Q_{\rm p}^{\prime} \gtsim 10^{7}$. Moreover, for initially large eccentricities (i.e., $e \gtsim 0.3$) and an initial $n/\Omega \approx 1$, the evolution of the system may be significantly affected and the infall of the planet  occurs as the orbital angular velocity at the periastron exceeds the stellar angular velocity. To prevent such an infall, we must assume that the star was initially rotating remarkably faster than considered above. Since the rotation  braking on the main sequence is expected to be small, this is hardly compatible with the present estimated  rotation period. We conclude, that the present eccentricity of CoRoT-11 is likely to be close to zero. In any case, in Sect.~\ref{present_diss}, we shall describe  observations that can constrain the present value of $e$ to allow us to discriminate among different scenarios for the system  evolution.

\subsubsection{Initial obliquity of the orbit}

\begin{figure}[t]
\centering
\includegraphics[width=8cm,height=6cm,angle=0]{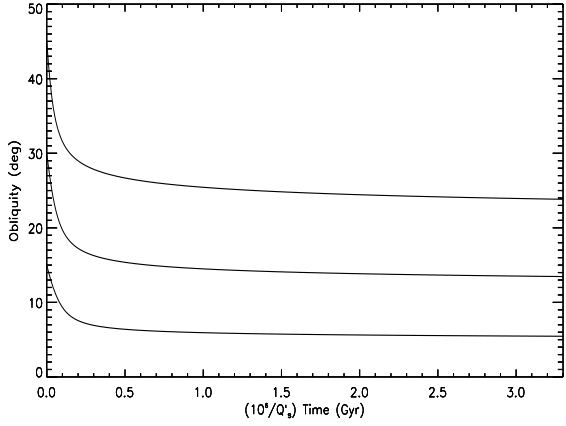} 
\caption{The forward evolution of the initial obliquity from the ZAMS  for different initial value of the angle itself. The planetary orbit is assumed circular and the initial orbital period is 0.7~days, while the initial rotation period of the star is 0.48~days for 
$\epsilon (t=0) = 45^{\circ}$, 0.60~days for $\epsilon (t=0) = 30^{\circ}$, and 0.67~days for $\epsilon (t=0) = 15^{\circ}$, to prevent the infall of the planet.  }
\label{obliquity_damping}
\end{figure}

The obliquity of the CoRoT-11 system may be significantly different from zero, as indeed found for several stars accompanied by hot Jupiters with $T_{\rm eff} > 6250$~K \citep{Winnetal10}. In the case of a system born close to $n / \Omega =1$, we need $\Omega \cos \epsilon > n$ to prevent the tidal infall of the planet and the observational result that CoRoT-11 is compatible with an aligned { system} is in agreement with this constraint. 
Initial values of the obliquity $\epsilon > 30^{\circ}$ require a fast-rotating star which is difficult to reconcile with the presently observed $ v \sin i$ and stellar radius. However, smaller values of the obliquity cannot be excluded. They are damped on a timescale that depends on the value of $Q_{\rm s}^{\prime}$, as illustrated in Fig.~\ref{obliquity_damping} where the time has been scaled inversely to the stellar tidal quality factor. These plots show that, after an initial remarkable damping of the obliquity, it asymptotically approaches a value about half its initial value that pratically corresponds to the present value. In Sect.~\ref{present_diss} we shall discuss how such a value can be constrained with accurate timing observations of the system. 

\section{Measuring the present tidal dissipation}
\label{present_diss}

The main conclusion of the previous section is that  the mean modified tidal quality factor of CoRoT-11a $Q_{\rm s}^{\prime} \ltsim 2 \times 10^{7}$, provided that the system started its evolution on the ZAMS from a nearly  synchronous state. Since $Q^{\prime}_{\rm s}$ is subject to remarkable variations during the evolution of the system, this result is of limited value to constrain dynamical tide theory, as in the case of the mean values of $Q_{\rm s}^{\prime}$ derived from close binaries in clusters. However, a much more interesting opportunity offered by CoRoT-11 is the possibility to measure the present value of $Q_{\rm s}^{\prime}$ from the rate of orbital period variation. Moreover, as we shall see, the rate of orbital period variation depends also on the orbit parameters, and we can extract a wealth of information from CoRoT-11 using the timing method.

\subsection{The circularized and aligned system hypothesis}
If the present orbit is circular, the semimajor axis is subject to a secular increase because the star is rotating faster than the orbital motion of the planet, so angular momentum is transferred from the stellar spin to the orbit via the tidal interaction. When $e=0$ and the stellar obliquity is negligible ($\epsilon \sim 0^{\circ}$), the rate of such a transfer depends on the system parameters, that are known to a good level of precision, and on $Q^{\prime}_{\rm s}$ only. Therefore, we can compute the rate of orbital period increase $dP_{\rm orb}/dt$ and use it to derive the difference $(O-C)$ between the observed time of mid-transit and that computed with a constant period ephemeris after $N$ orbital periods, viz.:
\begin{equation}
(O-C) (N) = \frac{1}{2} (\frac{dP_{\rm orb}}{dt})P_{\rm orb} N (N+1),
\label{o-c_dep}
\end{equation}
where we have assumed that the constant period of the ephemeris is the orbital period at $N=0$, i.e., at the epoch of the first transit. Using the tidal evolution model of \citet{Leconteetal10} and assuming
$e=0$, $\epsilon =0^{\circ}$, and the maximum stellar rotation period $P_{\rm rot}=1.73$~days, we find:
\begin{equation}
(O-C) = 8.34 \left(\frac{Q_{\rm s}^{\prime}}{10^{6}} \right)^{-1} \; \; \;  \mbox{s}
\label{o-c}
\end{equation}
after $N=3000$ orbital periods, i.e., 24.6 years. 

The  epoch of  mid transit can be determined with an accuracy that depends on the  depth and  duration of the transit, the accuracy  of the photometry, and the presence of  noise sources, e.g., related to stellar activity. In the case of CoRoT-11, since the star shows no signs of magnetic activity, the precision of the transit timing is dominated by the photometric accuracy. \citet{Gandolfietal10} report in their Table~4 a precision of $\sim 25$~s for the initial epoch of their ephemeris as obtained from a sequence of $\sim 50$ transits observed with a photometric accuracy of $\sim 250$ parts per million (ppm) and a cadence of $\sim 130$~s. 
Recently, \citet{Winnetal09a} using the Magellan/Baade 6.5-m telescope, observed two transits of the hot Jupiter WASP-4b, reaching a photometric accuracy of $\sim 500$ ppm (standard deviation) in their individual measurements obtained with a cadence of 30~s. The average standard deviation of the epoch of  mid-transit is $\sim 5$~s which can be assumed as the limit presently attainable with ground-based photometry. 
The transit of WASP-4b has a depth of $\approx 0.028$ mag and the star has a magnitude of $V \simeq 12.5$, while that of CoRoT-11 has a depth of $\approx 0.011$ mag and the star has $V = 12.94$. 
Therefore, it is reasonable to assume that a timing accuracy of $0.5-1$~s can be reached with a space-borne dedicated  telescope of aperture $2-3$~m, since the precision is expected to scale with the accuracy of the photometry that should reach $200-300$~ppm in $\sim 15$~s integration time. Such an instrument would allow us to detect the expected transit time variation (hereafter TTV) with  $3\sigma$ significance after an interval of $\sim 25$ years for $Q_{\rm s}^{\prime} \sim 5 \times 10^{6}$. Even in the case that no significant TTV is detected, a stringent lower limit  can be placed on  $Q^{\prime}_{\rm s}$, with the sensitivity increasing as the square of the  considered time baseline (cf. Eq.~\ref{o-c_dep}).

\subsection{The eccentric system hypothesis}
A significant eccentricity of the planetary orbit can affect the $O-C$ of the TTV, particularly when tidal dissipation in the planet is not negligible. This happens because the angular momentum of the planetary orbit is given by $L_{\rm p} = [(M_{\rm s} M_{\rm p}/(M_{\rm s} + M_{\rm p})] n a^{2} \sqrt{1- e^{2}}$, so  an increase of $L_{\rm p}$ does not necessarily lead to an increase of $a$ when the eccentricity is damped. 
Assuming $e=0.2$, $Q_{\rm s}^{\prime} = 5 \times 10^{6}$, and $Q_{\rm p}^{\prime}=10^{6}$, we find $O-C = -8.33$~s after 24.6 yr, owing to the strong dissipation inside the planet that dumps the eccentricity with a timescale of $\sim 120$~Myr. If the dissipation inside the planet is smaller, the absolute value of the $O-C$ 
decreases. For instance, if $Q_{\rm p}^{\prime}=10^{7}$, we obtain $O-C= 0.49$~s after 24.6 yr. 

Furthermore, if the orbit of CoRoT-11 is indeed eccentric, the measurement of the $O-C$ of the primary transit are not sufficient to determine $Q_{\rm s}^{\prime}$. This happens because in addition to the tidal evolution, also the precession of the line of the apsides (due to the planetary and stellar tidal deformations and general relativistic effects) produces a TTV \citep[see ][]{MiraldaEscude02,RagozzineWolf09}. For $e=0.2$, and a planet Love number $k_{2} = 0.3$, similar to that of Saturn, we have a precession period of only $\sim 3300$~yr, that leads to a maximum $O-C = 18.1$~s after $N=3000$ orbital periods, i.e., 24.6~yr. To separate the effects of the apsidal precession from those of tidal origin, we note that the apsidal precession produces an oscillation of the $O-C$ of the mid of the secondary eclipse, i.e., when the planet is occulted by the star, that is always shifted in phase by $180^{\circ}$ with respect to that of the primary transit. In other words, the two oscillations have always opposite signs as it is well known in the case of the apsidal precession of close binaries \citep[see, e.g., ][]{Gimenezetal87}. On the other hand, the $O-C$ of tidal origin has the same sign for both the primary transit and the secondary eclipse, so it can be separated from that arising from  apsidal precession.

 We conclude that a measurement of the eccentricity (or a determination of a more stringent upper limit) is critical to derive $Q_{\rm s}^{\prime}$ from the $O-C$ observations. A direct constraint on the eccentricity can be obtained from the observation of the time of the secondary eclipse.  \citet{Gandolfietal10} did not find evidence of a secondary eclipse in the CoRoT light curve which sets an upper limit of $\sim 100$~ppm to its depth in the spectral range $300-1100$ nm. However, it should be possible to detect the secondary eclipse in the infrared, thus deriving the value of $e \cos \omega$, where $\omega $ is the longitude of the periastron, from the deviation of the time of the secondary eclipse from half the orbital period. In the case of HD~189733, observing in a band centred at 8~$\mu$m with IRAC onboard of Spitzer, \citet{Knutsonetal07} found that the secondary eclipse occurred $120 \pm 24$~s after the time expected for a circular orbit, leading to $ e \cos \omega = 0.001 \pm 0.0002$. The same method has been applied to constrain the eccentricity of the orbit of CoRoT-2 finding $ e \cos \omega = -0.00291 \pm 0.00063$ \citep{Gillonetal10}. 
Moreover, the duration of the primary transit and the secondary eclipse can provide a measurement of $ e \sin \omega$  although with a precision significantly smaller than in the case of the secondary eclipse timing \citep[see Sect.~2.5 of][]{RagozzineWolf09}. In conclusion, we recommend to perform infrared photometry of CoRoT-11 to constrain its orbital eccentricity. 

We note that the present fast rotation of CoRoT-11a can in principle excite the eccentricity of the orbit during the phase of the evolution when $P_{\rm rot}/P_{\rm orb} < 11/18$ \citep[cf. Eq.~(6) of ][]{Leconteetal10}. However, the tidal dissipation inside the planet is  effective at damping the eccentricity and only for $Q^{\prime}_{\rm p} \geq 2.5 \times 10^{7}$ we find an increasing eccentricity for the present state of the system. { Imposing that the eccentricity is always damped along the past evolution of the system, we find $Q_{\rm p}^{\prime} \leq 1.6 \times 10^{7}$ for  $Q^{\prime}_{\rm s} \geq 4 \times 10^{6}$. }

\subsection{Non-negligible obliquity hypothesis}

If the present obliquity of the system is not negligible, the rate of orbital period increase and the corresponding $O-C$ in Eq.~(\ref{o-c}) must be multiplied by:
\begin{equation}
 \frac{\left( \frac{P_{\rm orb}}{P_{\rm rot}} \right) \cos \epsilon-1}{
\left( \frac{P_{\rm orb}}{P_{\rm rot}} \right) - 1},
\end{equation} 
 because the component of the stellar angular momentum interacting with the orbit is $L_{\rm s} \cos \epsilon $, where $L_{\rm s}$ is the modulus of the angular momentum of the star \citep[cf. Eq.~(2) of ][]{Leconteetal10}. 

It is interesting to note that the obliquity can be constrained not only through the observation of the Rossiter-McLaughlin effect, but also by measuring the precession of the orbital plane of the planet around the total angular momentum of the system that is expected when $\epsilon \not= 0^{\circ}$. Such a precession arises 
 because the stellar gravitational quadrupole moment is non-negligible due to its fast rotation. More precisely, adopting a stellar Love number  $k_{2 \rm s} = 0.0138$ after \citet{Claret95} and $P_{\rm rot} = 1.73$ days, we find $J_{2} = 4.19 \times 10^{-5}$. The contribution to $J_{2}$ arising from the tidal deformation of the star induced by the planet is about $1100$ times smaller and can be safely neglected. 

Even assuming an obliquity of a few degrees, the precession period is so short as to give measurable effects. For instance, for $\epsilon = 5^{\circ}$, the precession of the orbital plane has a period of $\sim 7.0 \times 10^{4}$ years, according to Eq.~(4) of \citet{MiraldaEscude02}. This produces a variation of the length of the transit chord across the disc of the star leading to a variation of the transit duration (transit duration variation, hereafter TDV). Applying Eq.~(12) of \citet{MiraldaEscude02} and assuming that the total angular momentum of the system is approximately perpendicular to the line of sight, we find a maximum TDV of $ \sim 8$~s in one year. Assuming a precision of $ \sim 40$~s in the measurement of the duration of the transit as obtained with CoRoT observations, such a variation should be easily detected in $\sim 5$~yr  and can provide a  lower limit for the stellar obliquity, remarkably better than achievable with the Rossiter-McLaughlin effect if $\epsilon \ltsim 15^{\circ}-20^{\circ}$.

\section{Role of a possible third body}

The use of TTV to measure the tidal dissipation inside the star (and the planet, in the case of an eccentric orbit), can be hampered by the presence of a distant third body  that induces apparent $O-C$ variations due to the light-time effect related to the revolution of the star-planet system around the common centre of mass. In principle, any light-time effect induced by a third body is periodic while the effects of tidal dissipation are secular, i.e., they do not change sign. However, if the orbital period of the third body around the barycentre of the system is longer than, say, three times the time baseline of our observations, it may become difficult or impossible to separate the two effects. 

To be specific, let us consider an $O-C$  of 3~s observed with a time baseline of $\Delta T \sim 25$~yr. If it is due to a light-time effect, the corresponding radial motion of the star-planet system is $\Delta z = c(O-C) \sim 10^{9}$~m, where $c$ is the speed of light, and the radial velocity of the barycentre of the system is $\Delta z / \Delta T \simeq 1.26$~m~s$^{-1}$, too small to be detectable with the present radial velocity accuracy \citep[cf. ][]{Gandolfietal10}. A third body moving on a coplanar circular orbit with a period of, say, $P_{\rm tb}=75$~yr, would have an orbital semimajor axis of $\sim 2.88 \times 10^{12}$~m and  a minimum mass of 0.2~M$_{\rm J}$ to induce a reflex motion of the above amplitude. Its direct detection is very difficult since its apparent separation from the central star  would be  only 0.034~arcsec, given an estimated distance to  CoRoT-11 of $560 \pm 30$~pc. 
Astrometric measurements with the GAIA or SIM satellites cannot be used to detect the reflex motion of CoRoT-11 since it would correspond to a maximum displacement of only $\sim 2.4$ $\mu$arcsec during an interval  of 5~yr, that is impossible to separate from the much larger proper motion of the system. 

If the orbit of the third body is inclined to the orbital plane of the star-planet system, the orbital plane of the planet will precede around the total angular momentum of the system. Assuming a third body mass of $0.2$~M$_{\rm J}$, a relative inclination of the orbits of $45^{\circ}$, and a semimajor axis of $\sim 2.88 \times 10^{12}$~m, equal to $\sim 20$~AU, Eq.~(8) of \citet{MiraldaEscude02} gives a precession period of $\sim 14$~Gyr, i.e., too long to be observable. 

Secular perturbations of the orbit of the planet by the distant third body can  be neglected because they   induce a variation of the $O-C$  of the order of $P_{\rm orb}/P_{\rm tb} \sim 10^{-4}$ of the light-time effect \citep[cf., e.g., ][]{Borkovitsetal03}. On the other hand, a perturber on a close-in orbit, possibly in resonance with the planet, would induce $O-C$ variations with a short periodicity and an amplitude up to a few minutes \citep{Agoletal05}, that would be easily  detectable.  

It is interesting to note that CoRoT-11a has a low level of magnetic activity in contrast to the case of stars with  lower effective temperature. This implies that the orbital period modulation induced by magnetic activity 
\citep[][]{Lanzaetal98,Lanza06} can be excluded as a source of $O-C$ variations in this case. This is not possible for cooler stars \citep{WatsonMarsh10} that are therefore less suitable to measure the TTV induced by tidal dissipation. 

\section{Conclusions}

{ Assuming an initial state of the CoRoT-11 system close to synchronization between the stellar spin and the orbital period of the planet, we can put constraints on the average modified tidal quality factor of its F-type star, finding $4 \times 10^{6} \ltsim Q_{\rm s}^{\prime} \ltsim 2 \times 10^{7}$.  Rigorously speaking, we should have expressed these constraints in terms of the average tidal time lag $\Delta t$ because our tidal equations are valid for a constant $\Delta t$ \citep[cf. Sect.~\ref{tidal_model} and ][]{Leconteetal10}. However, in view of the uncertainty on the above constraints, we prefer to give them in terms of $Q_{\rm s}^{\prime}$ that varies by a factor up to  $\approx 4-5$ during the evolution, if $\Delta t$ is assumed to be constant (cf. Sects.~\ref{tidal_model} and \ref{back_integ}). 

An initial state close to synchronization is not the only  possible one. However, the minimum estimated age of the system of $\sim 1$~Gyr implies $Q_{\rm s}^{\prime} \gtsim 4 \times 10^{6}$, otherwise the evolution from any initial state with $P_{\rm orb}> P_{\rm rot}$ would be too fast, leading to a present $P_{\rm orb}$ longer than observed. An upper limit on $Q_{\rm s}^{\prime}$ can be set when a larger statistics of the values of the $n/\Omega = P_{\rm rot}/P_{\rm orb}$ ratio will be available because for $Q_{\rm s}^{\prime} \gtsim 5 \times 10^{7}$ the initial value of $n/\Omega$ is predicted to be $\sim 0.25$ that is smaller than the presently observed minimum $n/\Omega \sim 0.5$ (see Sect.~\ref{back_integ}).

The above limits on $Q_{\rm s}^{\prime}$ are only a factor of $2-15$ smaller than the average value predicted by  \citet{BarkerOgilvie09} which can be considered a success of the dynamic tide theory of \citet{OgilvieLin07} given the current uncertainty on the tidal dissipation processes occurring in stellar convection zones. } 

We find that CoRoT-11 is one of the best candidates to look for orbital period variations related to tidal evolution by monitoring its transits and secondary eclipses in the optical and in the infrared passbands. In contrast to several systems hosting very hot Jupiters, e.g., WASP-12 or WASP-18, whose  orbital decay may be too slow to be measurable \citep{OgilvieLin07,Lietal10}, the parameters of CoRoT-11 appear to be ideal for a detection by measuring the times of mid-transit and mid-eclipse with an accuracy of at least $1-5$~s along a time baseline of $\sim 25$~yr. This is a consequence of the peculiar ratio of the rotation period of the star to the orbital period of the planet that is presently the smallest among systems having a radial velocity curve compatible with zero eccentricity. { The test is particularly sensitive to values of $Q_{\rm s}^{\prime}$ between $10^{5}$ and $10^{6}$ leading to a fast tidal evolution of the system, as discussed in Sect.~\ref{back_integ}. }

The CoRoT-11 system is also particular because the precession of the orbital plane due to a non-zero obliquity of the  stellar spin can be measured through the variation of the duration of the transit on a timescale as short as $5-10$~yr by means of ground-based observations with telescopes of the $6-8$~m class. Note that even an obliquity  as small as $5^{\circ}$ can be  detected with this method. On the other hand,  the eccentricity of the planetary orbit can be well constrained by measuring the times of the secondary eclipses in the infrared and their duration. If the system turns out to have an eccentric orbit, this puts a constraint also on the minimum value of the planet quality factor $Q_{\rm p}^{\prime}$ if we assume that the eccentricity is of primordial origin, or is a clear indication of the presence of a perturbing body that excites it \citep[e.g., ][]{TakedaRasio05}. The light-time effect due to a distant third body is indeed the only phenomenon that can seriously hamper the detection of the transit time variations expected from the tidal orbital evolution. The detection of a distant companion may be particularly difficult, but this limitation is  present also in any other system candidate for a direct measurement of the tidal dissipation.

\begin{acknowledgements}
The authors are grateful to the Referee, Prof.~J.-P.~Zahn, and to the Editor, Dr. T.~Guillot, for several valuable comments on the first version of the manuscript, and to Drs. L.~Carone and S.~Ferraz-Mello for interesting discussions. Active star research and exoplanetary studies at INAF-Catania Astrophysical Observatory and the Department of Physics and Astronomy of Catania University are funded by MIUR ({\it Ministero dell'Istruzione, Universit\`a e Ricerca}) and by {\it Regione Siciliana}, whose financial support is gratefully
acknowledged. 
This research  made use of the ADS-CDS databases, operated at the CDS, Strasbourg, France.
\end{acknowledgements}

\begin{footnotesize}

\end{footnotesize}



\begin{thebibliography}{}
\bibliographystyle{aa}

\bibitem[Agol et al. (2005)]{Agoletal05} 
Agol, E., Steffen, J., Sari, R., \& Clarkson, W.\ 2005, \mnras, 359, 567

\bibitem[Ammler-von Eiff et al. (2009)]{Ammleretal09}
Ammler-von Eiff, M., Santos, N. C., Sousa, S. G., et al.\ 2009, \aap, 507, 523

\bibitem[Anderson et al. (2010)]{Andersonetal10}
{Anderson}, D.~R.,{Hellier}, C., {Gillon}, M., et al.\ 2010, \apj, 709, 159

\bibitem[Bakos et al, (2007)]{Bakosetal07}
Bakos, G.~{\'A}., {Kov{\'a}cs}, G., Torres, G., et al.\ 2007, \apj, 670, 826

\bibitem[Barker \& Ogilvie (2009)]{BarkerOgilvie09}
Barker, A.~J., \& Ogilvie, G.~I.\ 2009, \mnras, 395, 2268 

\bibitem[Borkovits et al. (2003)]{Borkovitsetal03} 
Borkovits, T., {\'E}rdi, B., Forg{\'a}cs-Dajka, E., \& Kov{\'a}cs, T.\ 2003, \aap, 398, 1091 

\bibitem[Bouchy et al. (2011)]{Bouchyetal11} 
{ Bouchy, F., et al.\ 2011, \aap, 525, A68 }


\bibitem[Bouvier et al. (1997)]{Bouvieretal97} 
{ Bouvier, J., Forestini, M., \& Allain, S.\ 1997, \aap, 326, 1023 }


\bibitem[Cameron et al. (2010)]{Cameronetal10}
{Cameron}, A.~C., {Guenther}, E., {Smalley}, B., et al.\ 2010, \mnras, 407, 507

\bibitem[Carone \& P\"atzold (2007)]{CaronePatzold07} 
{Carone, L., P\"atzold, M.\ 2007, \planss, 55, 643 }

\bibitem[Claret (1995)]{Claret95}
Claret, A.\ 1995, \aaps, 109, 441 

\bibitem[Cohen et al.(2010)]{Cohenetal10}
Cohen, O., Drake, J.~J, Kashyap, V.~L., Sokolov, I.~V., \& Gombosi, T.~I.\ 2010, \apjl, 723, L64

\bibitem[Gandolfi et al. (2010)]{Gandolfietal10} 
Gandolfi, D., et al. 2010, \aap, 524, A55

\bibitem[Gibson et al. (2008)]{Gibsonetal08}
Gibson, N. P., Pollacco, D., Simpson, E. K., et al. 2008, \aap, 492, 603

\bibitem[Gillon et al. (2010)]{Gillonetal10} 
Gillon, M., et al.\ 2010, \aap, 511, A3 

\bibitem[Gimenez et al. (1987)]{Gimenezetal87}
Gimenez, A., Kim, C.-H., \& Nha, I.-S.\ 1987, \mnras, 224, 543

\bibitem[Goodman \& Lackner (2009)]{GoodmanLackner09}
Goodman, J., \& Lackner, C.\ 2009, \apj, 696, 2054

\bibitem[Hebb et al. (2009)]{Hebbetal09}
Hebb, L., Cameron, A. C., Loeillet, B., et al. 2009, \apj, 693, 1920

\bibitem[Hellier et al. (2009a)]{hellieretal09a}
Hellier, C., Anderson, D., Gillon, M., et al.\ 2009, \apjl, 690 , L89

\bibitem[Hellier at al. (2009b)]{hellieretal09b}
{Hellier}, C., {Anderson}, D.~R. , {Cameron}, A.~C., et al.\ 2009, \nat, 460, 1098

\bibitem[Hut(1980)]{Hut80}
Hut, P.\ 1980, \aap, 92, 167 

\bibitem[Hut (1981)]{Hut81}
Hut, P.\ 1981, \aap, 99, 126

\bibitem[Irwin \& Bouvier (2009)]{IrwinBouvier09} 
{ Irwin, J., \& Bouvier, J.\ 2009, IAU Symposium, 258, 363 }


\bibitem[Jenkins et al. (2010)]{Jenkinsetal10}
Jenkins, J. M., Borucki, W. J., Koch, D. G., et al.\ 2010, \apj, 724, 1108

\bibitem[Johns-Krull et al. (2008)]{JohnsKrulletal08}
Johns-Krull, C. M., McCullough, P. R., Burke, C. J., et al. 2008, \apj, 677, 657

\bibitem[Johnson et al. (2009)]{Johnsonetal09}
Johnson, J., Winn, J., Albrecht, S., et al.\ 2009, \pasp , 121 , 1104

\bibitem[Joshi et al. (2008)]{Joshietal08}
Joshi, Y., Pollacco, D., Cameron, A. C., et al.\ 2008, \mnras, 392 , 1532

\bibitem[Kipping et al (2010)]{Kippingetal10}
{Kipping}, D.~M., {Bakos}, G.~{\'A}., {Hartman}, J., et al.\ 2010, \apj, 725, 2017

\bibitem[Koch et al. (2010)]{Kochetal10}
{Koch}, D.~G., {Borucki}, W.~J., {Rowe}, J.~F., et al.\ 2010, \apjl, 713, L131

\bibitem[Knutson et al. (2007)]{Knutsonetal07}
Knutson, H.~A., et al.\ 2007, \nat, 447, 183

\bibitem[Lainey et al. (2009)]{Laineyetal09} 
Lainey, V., Arlot, J.-E., Karatekin, {\"O}., \& van Hoolst, T.\ 2009, \nat, 459, 957 

\bibitem[Lanza(2006)]{Lanza06}
Lanza, A.~F.\ 2006, \mnras, 369, 1773

\bibitem[Lanza (2010)]{Lanza10}
Lanza, A. F.\ 2010, \aap, 512, A77

\bibitem[Lanza et al.(1998)]{Lanzaetal98}
Lanza, A.~F., Rodono, M., \& Rosner, R.\ 1998, \mnras, 296, 893

\bibitem[Leconte et al. (2010)]{Leconteetal10}
Leconte, J., Chabrier, G., Baraffe, I., \& Levrard, B.\ 2010, \aap, 516, A64 

\bibitem[Levrard et al. (2009)]{Levrardetal09} 
{ Levrard, B., Winisdoerffer, C., \& Chabrier, G.\ 2009, \apjl, 692, L9 } 

\bibitem[Li et al.(2010)]{Lietal10} 
Li, S.-L., Miller, N., Lin, D.~N.~C., \& Fortney, J.~J.\ 2010, \nat, 463, 1054 

\bibitem[Lin et al.(1996)]{Linetal96}
Lin, D.~N.~C., Bodenheimer, P., \& Richardson, D.~C.\ 1996, \nat, 380, 606 

\bibitem[Loeillet et al. (2008)]{Loeilletetal08}
Loeillet, B., Shporer, A., Bouchy, F., et al.\ 2008, \aap, 481, 529

\bibitem[Lovelace et al. (2008)]{Lovelaceetal08}
Lovelace, R.~V.~E., Romanova, M.~M., \& Barnard, A.~W.\ 2008, \mnras, 389, 1233 

\bibitem[Mardling \& Lin (2002)]{MardlingLin02}
Mardling, R.~A., \& Lin, D.~N.~C.\ 2002, \apj, 573, 829

\bibitem[Matsumura et al. (2008)]{Matsumuraetal08} 
Matsumura, S., Takeda, G., \& Rasio, F.~A.\ 2008, \apjl, 686, L29 

\bibitem[McCullough et al. (2008)]{McCulloughetal08}
{McCullough}, P.~R., {Burke}, C.~J. , {Valenti}, J.~A., et al.\ 2008, submitted to \apj, {\tt arXiv:0805.2921v1}

\bibitem[Miralda-Escud{\'e}(2002)]{MiraldaEscude02}
Miralda-Escud{\'e}, J.\ 2002, \apj, 564, 1019

\bibitem[Noyes et al. (2008)]{Noyesetal08}
{Noyes}, R.~W., Bakos, G.~{\'A}., {Torres}, G., et al.\ 2008, \apjl, 673,L79

\bibitem[Narita et al. (2009)]{Naritaetal09}
Narita, N., Sato, B., Hirano, T., \& Tamura, M.\ 2009 \pasj , 61 , L35

\bibitem[Narita et al. (2010)]{Naritaetal10}
{Narita}, N., {Hirano}, T., {Sanchis-Ojeda}, R., et al.\ 2010, \pasj, 62, L61

\bibitem[Ogilvie \& Lin(2004)]{OgilvieLin04}
Ogilvie, G.~I., \& Lin, D.~N.~C.\ 2004, \apj, 610, 477

\bibitem[Ogilvie \& Lin (2007)]{OgilvieLin07}
Ogilvie, G.~I., \& Lin, D.~N.~C.\ 2007, \apj, 661, 1180 

\bibitem[Ohta et al. (2005)]{Ohtaetal05} 
Ohta, Y., Taruya, A., \& Suto, Y.\ 2005, \apj, 622, 1118 


\bibitem[Pollacco et al. (2008)]{Pollaccoetal08}
Pollacco, D., Skillen, I., Cameron, C. A., et al. 2008, \mnras, 385, 1576

\bibitem[Ragozzine \& Wolf(2009)]{RagozzineWolf09}
Ragozzine, D., \& Wolf, A.~S.\ 2009, \apj, 698, 1778

\bibitem[Sari \& Goldreich (2004)]{SariGoldreich04} 
Sari, R., \& Goldreich, P.\ 2004, \apjl, 606, L77 

\bibitem[Shporer et al, (2009)]{Shporeretal09}
Shporer, A., Bakos, G.~{\'A}., Bouchy, F., et al. 2009, \apj, 690, 1393

\bibitem[Siess et al. (2000)]{Siessetal00} 
Siess, L., Dufour, E., \& Forestini, M.\ 2000, \aap, 358, 593 


\bibitem[Simpson et al. (2010)]{Simpsonetal10}
Simpson, E., Barros, S., Brown, A., et al.\ 2010, submitted to \aj, {\tt arXiv:1009.3470v1}

\bibitem[Snellen et al. (2009)]{Snellenetal09}
Snellen, I. A. G., Koppenhoefer, J., van der Burg, R. F. J., et al. 2009, \aap, 497, 545

\bibitem[Takeda \& Rasio (2005)]{TakedaRasio05} 
Takeda, G., \& Rasio, F.~A.\ 2005, \apj, 627, 1001 

\bibitem[Torres et al. (2008)]{Torresetal08}
Torres, G., Winn, J., \& Holman, M.\ 2008, \apj, 677, 1324

\bibitem[Torres et al. (2010)]{Torresetal10}
Torres, G., Bakos,  G.~{\'A}., Hartman J., et al.\ 2010, \apj, 715 , 458

\bibitem[Vidotto et al. (2010)]{Vidottoetal10}
Vidotto, A.~A., Opher, M., Jatenco-Pereira, V., \& Gombosi, T.~I.\ 2010, \apj, 720, 1262 

\bibitem[Watson \& Marsh (2010)]{WatsonMarsh10} 
Watson, C.~A., \& Marsh, T.~R.\ 2010, \mnras, 405, 2037

\bibitem[West et al. (2009)]{Wessetal09}
{West}, R.~G., {Anderson}, D.~R., {Gillon}, M. et al.\ 2009, \aj, 137, 4834

\bibitem[Winn et al.(2009)]{Winnetal09a}
Winn, J.~N., Holman, M.~J., Carter, J.~A. et al.\ 2009a, \aj, 137, 3826

\bibitem[Winn et al. (2009)]{Winnetal09b}
Winn, J. N., Johnson, J. A., Fabrycky, D., et al. 2009b, \apj, 700, 302

\bibitem[Winn et al. (2010)]{Winnetal10} 
Winn, J.~N., Fabrycky, D.,  Albrecht, S., \& Johnson, J.~A.\ 2010, \apjl, 718, L145 

\bibitem[Wolff \& Simon (1997)]{WolffSimon97} 
Wolff, S., \& Simon, T.\ 1997, \pasp, 109, 759 

\bibitem[Zahn(1977)]{Zahn77}
Zahn, J.-P.\ 1977, \aap, 57, 383 

\bibitem[Zahn(1989)]{Zahn89}
Zahn, J.-P.\ 1989, \aap, 220, 112 

\bibitem[Zahn(2008)]{Zahn08}
Zahn, J.-P.\ 2008, EAS Publications Series, 29, 67

\end{thebibliography}
\end{document}